\documentclass[aps,prb,showpacs,longbibliography,floatfix,reprint]{revtex4-1}

\usepackage{graphicx}
\usepackage{subfigure}
\usepackage{amsfonts}
\usepackage{amsmath}
\usepackage{amssymb}

\begin{document}

\title{Decoherence in current induced forces: Application to adiabatic quantum motors.}
\author{Lucas J. Fern\'andez-Alc\'azar$^{1}$}
\affiliation{$^{1}$Instituto de F\'{\i}sica Enrique Gaviola and Facultad de Matem\'{a}tica 
Astronom\'{\i}a y F\'{\i}sica, Universidad Nacional de C\'{o}rdoba,
Ciudad Universitaria, C\'{o}rdoba, 5000, Argentina.} 
\affiliation{$^{2}$Facultad de Ciencias Qu\'{\i}micas, Universidad Nacional de C\'{o}rdoba,
Ciudad Universitaria, C\'{o}rdoba, 5000, Argentina.}

\author{Ra\'ul A. Bustos-Mar\'un$^{1,2}$}
\thanks{Corresponding author: \texttt{rbustos@famaf.unc.edu.ar}}
\affiliation{$^{1}$Instituto de F\'{\i}sica Enrique Gaviola and Facultad de Matem\'{a}tica 
Astronom\'{\i}a y F\'{\i}sica, Universidad Nacional de C\'{o}rdoba,
Ciudad Universitaria, C\'{o}rdoba, 5000, Argentina.} 
\affiliation{$^{2}$Facultad de Ciencias Qu\'{\i}micas, Universidad Nacional de C\'{o}rdoba,
Ciudad Universitaria, C\'{o}rdoba, 5000, Argentina.}

\author{Horacio M. Pastawski$^{1}$}
\affiliation{$^{1}$Instituto de F\'{\i}sica Enrique Gaviola and Facultad de Matem\'{a}tica 
Astronom\'{\i}a y F\'{\i}sica, Universidad Nacional de C\'{o}rdoba,
Ciudad Universitaria, C\'{o}rdoba, 5000, Argentina.} 
\affiliation{$^{2}$Facultad de Ciencias Qu\'{\i}micas, Universidad Nacional de C\'{o}rdoba,
Ciudad Universitaria, C\'{o}rdoba, 5000, Argentina.}

\pacs{85.85.+j, 73.23.-b, 03.65.Yz.}

\begin{abstract}
Current induced forces are not only related with the discrete nature of electrons but also with its quantum character. It is natural then to wonder about the effect of decoherence.
Here, we develop the theory of current induced forces including dephasing processes and we apply it to study adiabatic quantum motors (AQMs). The theory is based on B\"uttiker's fictitious probe model which here is reformulated for this particular case. We prove that it accomplishes fluctuation-dissipation theorem. We also show that, in spite of decoherence, the total work performed by the current induced forces remains equal to the pumped charge per cycle times the voltage.
We find that decoherence affects not only the current induced forces of the system but also its intrinsic friction and noise, modifying in a non trivial way the efficiency of AQMs. 
We apply the theory to study an AQM inspired by a classical peristaltic pump where we surprisingly find that decoherence can play a crucial role by triggering its operation. Our results can help to understand how environmentally induced dephasing affects the quantum behavior of nano-mechanical devices.
\end{abstract}

\maketitle

\section{Introduction.}

Nano-mechanical devices in general and nanomotors in particular are topics that have attracted much attention in recent years.\cite{RefGen1,RefGen2,RefGen3,RefGen4,RefGen5,RefGen6}
The working principle of most of these experimental and theoretical proposals rely on classical physics.
However, at nanoscale one may benefit from the many phenomena emerging from quantum interferences.
In this direction, reverse quantum pumping has been recently proposed as the basic mechanism by which DC-current induced forces can drive a nanoscopic motor.
Such a device is now known as ``Adiabatic Quantum Motor'' (AQM).\cite{AQM13,Bailey,Dundas,Zhang}.
Despite the classical description of the motor's movement, one can profit from the quantum nature of current induced forces. 
Indeed, one can boost the efficiency of AQMs by exploiting the  interferences present in a Thouless pump \cite{AQM13}.

Many natural questions arise when one addresses current induced forces in nano-mechanical devices. Will their quantum behavior survive under realistic non-ideal situations? Which is the effect of environmentally induced decoherence and how can we model it?  Will the effect of decoherence be always counter-productive for an AQM?
Solving these issues will provide a better understanding of their working principles and ways to assess the feasibility of their  experimental implementations.

In this work, the theory of current induced forces based on scattering matrices \cite{vonOppen0,vonOppen,vonOppen12} is extended to include decoherent events. Our approach is based on a reformulation of B\"{u}ttiker's fictitious probe model \cite{Bt86PRB,DP90,GLBE2}. This allows us to address the effect of decoherence on non-equilibrium current induced forces, friction coefficients, and fluctuating forces with a focus on AQMs.

\section{Theory.}

\subsection{Langevin equation and current induced forces.}

In the non-equilibrium Born-Oppenheimer approximation, the dynamical degrees of freedom of a system are slow as compared to the electron dynamics. In this limit, we can treat the mechanical degrees of freedom as a classical field acting on the electrons. Since, the movement of a motor is cyclic, one can reduce its many degrees of freedom to a single rotational coordinate, $x$. Hence, one can describe the rotor's dynamics by the 1-D Langevin equation,
\begin{equation}
M\ddot{x}+\frac{\mathrm{d}U}{\mathrm{d}x}=F-\gamma \dot{x}+\xi ,\label{eq:Langevin}
\end{equation}
where $M$ is the mass of the rotor (or the moment of inertia),
and $U$ is some classical external potential that can also be introduced.
The right hand side of Eq. \ref{eq:Langevin} accounts for the current induced forces where $F$ is
a mean adiabatic reaction force. The second term is a friction (dissipative) force where $\gamma $ is the
friction coefficient. The last term, $\xi $, accounts for the force fluctuations.
These current induced forces have a quantum origin and practical
expressions in terms of the scattering matrices are given in Refs.
\cite{vonOppen0,vonOppen,vonOppen12} which are our starting point.

Like in Brownian motion, the interaction of the electrons with the rotor gives rise to a dissipation mechanism and a fluctuating force. At equilibrium, the only non-vanishing contribution to the friction coefficient is the symmetric contribution $\gamma ^{s,eq}$.\cite{vonOppen0,vonOppen,vonOppen12}
On the other hand, there is a fluctuating force $\xi (t)$ whose self-correlation $D$, defined by $\left\langle \xi (t)\xi (t^{\prime })\right\rangle \approx D\delta (t-t^{\prime })$, is assumed locally correlated in time. These quantities must be related by the \textit{fluctuation-dissipation theorem} (FDT), 
i.e., they satisfy $D=2K_{B}T~\gamma ^{s,eq}$, where $K_{B}T$ is the thermal energy.
We will consider the forces only up first order in $e\texttt{V}$ and/or $\dot{x}$. That implies that only the equilibrium contributions to the friction coefficient and $D$ will be taken into account.

\subsection{Decoherence and null current condition.}

Decoherence is included by connecting the system to a fictitious voltage probe $\phi$. The voltmeter condition imposes $I_{\phi }=0$, which provides charge conservation, while the reservoir character of the fictitious probe ensures the loss of memory of the re-injected electrons. Strictly speaking, this also involves inelastic events that redistribute electron's energy.
However, in a linear response, $e\texttt{V}\rightarrow 0$, inelasticity is just reduced to a stochastic dephasing of the wave function, i.e. decoherence.\cite{Lucas}

The total current flowing at lead $m$ is
\begin{equation}
I_{m}=I_{m}^{Bias}+I_{m}^{Pump}+\delta I_{m}, \label{eq:I}
\end{equation}
where $I_{m}^{Bias}$ is the non-equilibrium current caused by an infinitesimal bias $e\texttt{V}$, $I_{m}^{Pump}$ is the pumped current due to a variation of $x$, and $\delta I_{m}$ accounts for current fluctuations.
Fictitious volage probe metod has proved useful to address a quite related problem, the quantum pumping with dephasing.\cite{CrBrow02,MkBt01}
While there, fluctuations could be disregarded, here, we need to consider them explicitly as they affect the dynamics of the system.

The non-equilibrium current incoming
to the system through the lead $m=L,\phi $, in a linear response limit and at low temperatures, is 
$I_{m}^{Bias}=(e/2\pi \hbar )\left( \sum_{n\neq m}T_{nm}\delta \mu
_{m}-\sum_{n\neq m}T_{nm}\delta \mu _{n}\right) $. Here, $T_{nm}$ is the
transmittance between leads $m$ and $n=L,R,\phi $; and $\delta \mu _{m}=\mu
_{m}-\mu _{R}$ is the chemical potential of $m$, taking $\mu _{R}$ as reference.

The pumped current $I_{m}^{Pump}$ through a lead $m$ is $I_{m}^{Pump}=e(
\mathrm{d}n_{m}/\mathrm{d}x)\dot{x}$, were $e$ is the electron charge, $\dot{
x}$ the velocity of the rotational parameter $x$, and the emissivity of the
lead $m$ is
\begin{equation}
\frac{\mathrm{d}n_{m}}{\mathrm{d}x}=\int \frac{\mathrm{d}\varepsilon }{2\pi 
\mathrm{i}}\left( -\frac{\partial f}{\partial \varepsilon }\right)
\mathrm{Tr}\left\{ \Pi _{m}
\frac{\mathrm{d}S}{\mathrm{d}x} S^{\dagger }
\right\} , \label{eq:emmissivity}
\end{equation}
where $f$ is the Fermi distribution, $\Pi _{m}$ is a projector onto the lead $m$, and $S$ is the scattering matrix. \cite{Brower98,MkBt01} 
By integration of $I_{L}^{Pump}$ we obtain the pumped charge per cycle through lead $L$,  \cite{CrBrow02,MkBt01}
\begin{equation}
Q_{L}^{}=e\oint \left[ \frac{\mathrm{d}n_{L}}{\mathrm{d}x}+\frac{
T_{L\phi }}{T_{L\phi }+T_{\phi R}}\frac{\mathrm{d}n_{\phi }}{\mathrm{d}x}
\right] \delta x.  \label{eq:QpumpXcycle}
\end{equation}

Current fluctuations contain both the non-equilibrium shot-noise and the
thermal noise. Both of them satisfy $\left\langle \delta I_{m}\right\rangle =0$,
and $\sum_{m}\delta I_{m}=0$. However, at equilibrium, the shot noise vanishes
and only survives thermal noise. 

The null current condition $I_{\phi }=0$ at lead $\phi$ directly imposes a condition to $\delta \mu _{\phi }$.
The value of $\delta \mu _{\phi }$ that ensures current cancellation of Eq. \ref{eq:I} is
\begin{equation}
\delta \mu _{\phi }=\frac{1}{T_{L\phi }+T_{\phi R}}\left(T_{L\phi}\delta \mu
_{L}-2\pi \hbar \frac{\mathrm{d}n_{\phi}}{\mathrm{d}x} \dot{x}
-\frac{2\pi \hbar }{e}\delta I_{\phi }\right).  \label{eq:deltaMuPhi}
\end{equation}
Since the variation of $x$ is slow respect to the electronic dynamics, we can consider that $\delta \mu _{\phi }$ adapts instantaneously to satisfy $I_\phi=0$ at all time.
  
\subsection{Current induced forces in presence of decoherence.}

By considering an infinitesimal bias, $\delta \mu _{L}=e\texttt{V}$, we can split the Fermi function into an equilibrium and an out-of-equilibrium contributions, $f_{m}=f_{0}+\Delta f_{m }$.
Then, the force generated by the
electric current, \cite{vonOppen0,vonOppen,vonOppen12} 
\begin{equation}
F=\sum_{m }\int \frac{\mathrm{d}\varepsilon }{2\pi \mathrm{i}}f_{m
}\mathrm{Tr}\left( \Pi _{m }S^{\dagger }\frac{\mathrm{d}S}{\mathrm{d}x}\right) ,
\label{eq:Force1}
\end{equation}
can be split as $F=F^{eq}+\Delta F$. The equilibrium force, $F^{eq}$, is
conservative and thus, it does not produce work. At low temperatures and for small $e\texttt{V}$, $\int
(\cdot )\Delta f_{m }\mathrm{d}\varepsilon \simeq (\cdot )\delta \mu
_{m }$. Using Eqs. \ref{eq:emmissivity}, \ref{eq:deltaMuPhi}, and \ref{eq:Force1} we calculate the non-conservative forces for systems without magnetic fields up to first order in $e\texttt{V}$ and $\dot{x}$,
\begin{eqnarray}
\Delta F &=&\left[ \frac{\mathrm{d}n_{L}}{\mathrm{d}x}+\frac{T_{L\phi }}{
T_{L\phi }+T_{\phi R}}\frac{\mathrm{d}n_{\phi }}{\mathrm{d}x}\right] e
\texttt{V}  \notag \\
&&-2\pi \hbar \frac{\dot{x}}{T_{L\phi }+T_{\phi R}}\left( \frac{\mathrm{d}
n_{\phi }}{\mathrm{d}x}\right) ^{2}  \label{eq:force} \\
&&+\frac{2\pi \hbar }{e}\frac{\delta I_{\phi }}{T_{L\phi }+T_{\phi R}}\frac{
\mathrm{d}n_{\phi }}{\mathrm{d}x}. \notag
\end{eqnarray}
The first term on the right hand side (RHS) of Eq. \ref{eq:force} is the non-equilibrium
force, $F^{ne}$, whose second term within the brackets is the decoherent contribution. This force is responsible for the work of the system, which is obtained by evaluating $W=\oint
\left( F^{eq}+F^{ne}\right) dx$, yielding
\begin{equation}
W=\oint dx\left[ \frac{\mathrm{d}n_{L}}{\mathrm{d}x}+\frac{T_{L\phi }}{
T_{L\phi }+T_{\phi R}}\frac{\mathrm{d}n_{\phi }}{\mathrm{d}x}\right] e\texttt{V}.
\label{eq:work}
\end{equation}
Comparing Eqs. \ref{eq:work} and \ref{eq:QpumpXcycle} we realize that the
work in presence of decoherence is proportional to the pumped charge per cycle times the voltage's bias,
\begin{equation}
W=Q \texttt{V}.  \label{eq:W=QV}
\end{equation}
This relation, proved valid for coherent AQMs, can be thought as a signature of the Onsager's reciprocal relations,\cite{Cohen} showing that the model is well behaved.

The second term on the RHS of Eq. \ref{eq:force} is a force proportional to the
velocity. Thus, one can associate it with a dissipative force whose origin is purely due to decoherence. The resulting friction coefficient $\gamma ^{\phi}$ is
\begin{equation}
\gamma ^{\phi }=\frac{2\pi \hbar }{T_{L\phi }+T_{\phi R}}\left( \frac{
\mathrm{d}n_{\phi }}{\mathrm{d}x}\right)^{2}. \label{eq:gammaPhi}
\end{equation}
This prove that decoherence inside the sample enables energy dissipation through an additional friction of the rotor.\cite{comment,comment2}

The third term on the RHS of Eq. \ref{eq:force} accounts for the fluctuations on the force induced by 
decoherence. The self-correlation of the fluctuating force $\xi _{\phi }$,
can be defined as $\left\langle \xi _{\phi }(t) \xi _{\phi }(t') \right\rangle \approx D_{\phi } \delta(t-t')$. 
Thus,
\begin{equation}
D_{\phi }=\left( \frac{2\pi \hbar }{e}\frac{1}{T_{L\phi }+T_{\phi R}}
\frac{\mathrm{d}n_{\phi }} {\mathrm{d}x} \right)^{2} S_{\phi} ,   
\label{eq:fluc_force_phi}
\end{equation}
where $S_{\phi }$, the spectrum power of the fluctuating current at the lead $\phi$, 
is defined as $\langle \delta I_{\phi}(t)I_{\phi}(t') \rangle \approx S_{\phi } \delta(t-t')$. 
In deducing Eq. \ref{eq:fluc_force_phi}, we use that transmittances and emissivities are already
mean values.
The only contribution to the current fluctuation which is
non-vanishing at equilibrium is the thermal fluctuation or
Nyquist-Johnson noise. \cite{review_Buttiker_noise} Thus, $S_{\phi}$ is characterized by 
\begin{equation}
S_{\phi} =2K_{B}T~\frac{e^{2}}{2\pi
\hbar }\left( T_{L\phi }+T_{\phi R}\right) .
\label{eq:fluc_current}
\end{equation}
Replacing Eq. \ref{eq:fluc_current} into Eq. \ref{eq:fluc_force_phi} yields
\begin{equation}
D_{\phi }=2K_{B}T~\gamma ^{\phi },  \label{eq:FDT_dec}
\end{equation}
which demonstrates that decoherent friction and fluctuating forces induced by decoherence are related by the FDT in our model as they should be.
\begin{figure}
\begin{center}
\includegraphics[width=2.2 in]{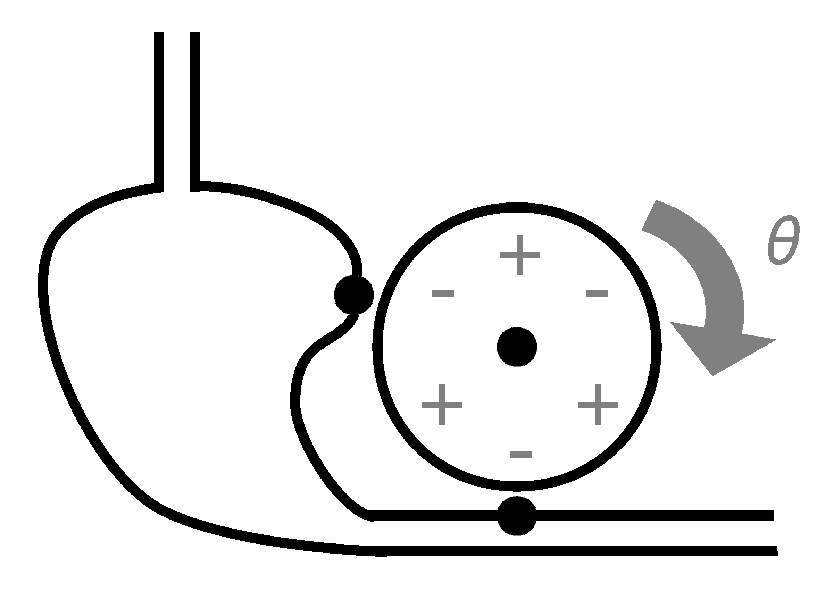}
\end{center}
\caption{Scheme of the type of system studied. A rotational device interacting with a quantum dot. Rotation of the motor changes the dot's energy as well as its coupling to one lead.}
\label{fig:scheme}
\end{figure}
\begin{figure}
     \begin{center}
      \subfigure{\includegraphics[width=2.5in, trim=0.5in 0.1in 0.1in 0.2in,clip=true]{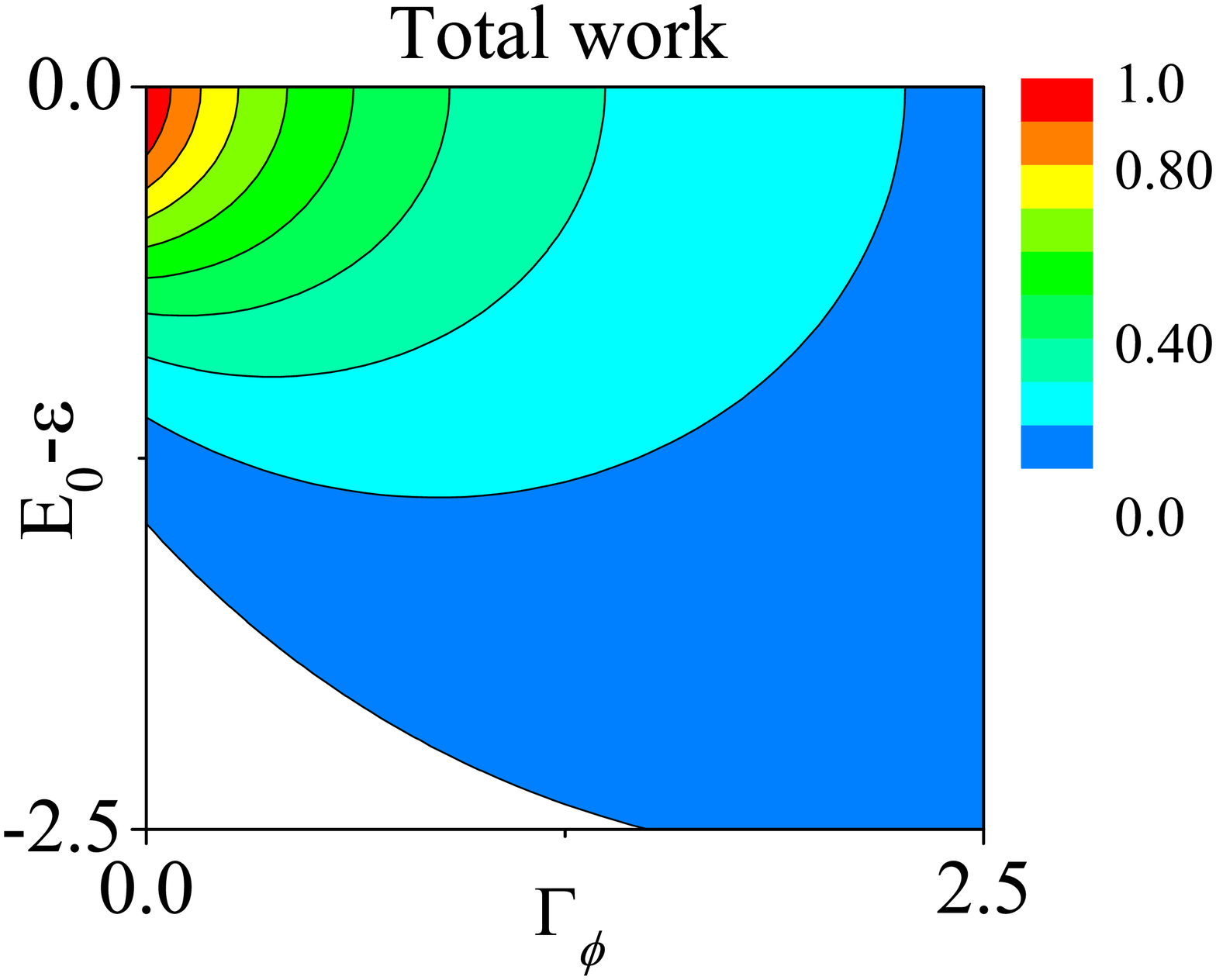}}        
                \\
      \subfigure{\includegraphics[width=1.5 in,height=1.5 in,trim=0.5in 0.1in 0.1in 0.2in,clip=true]{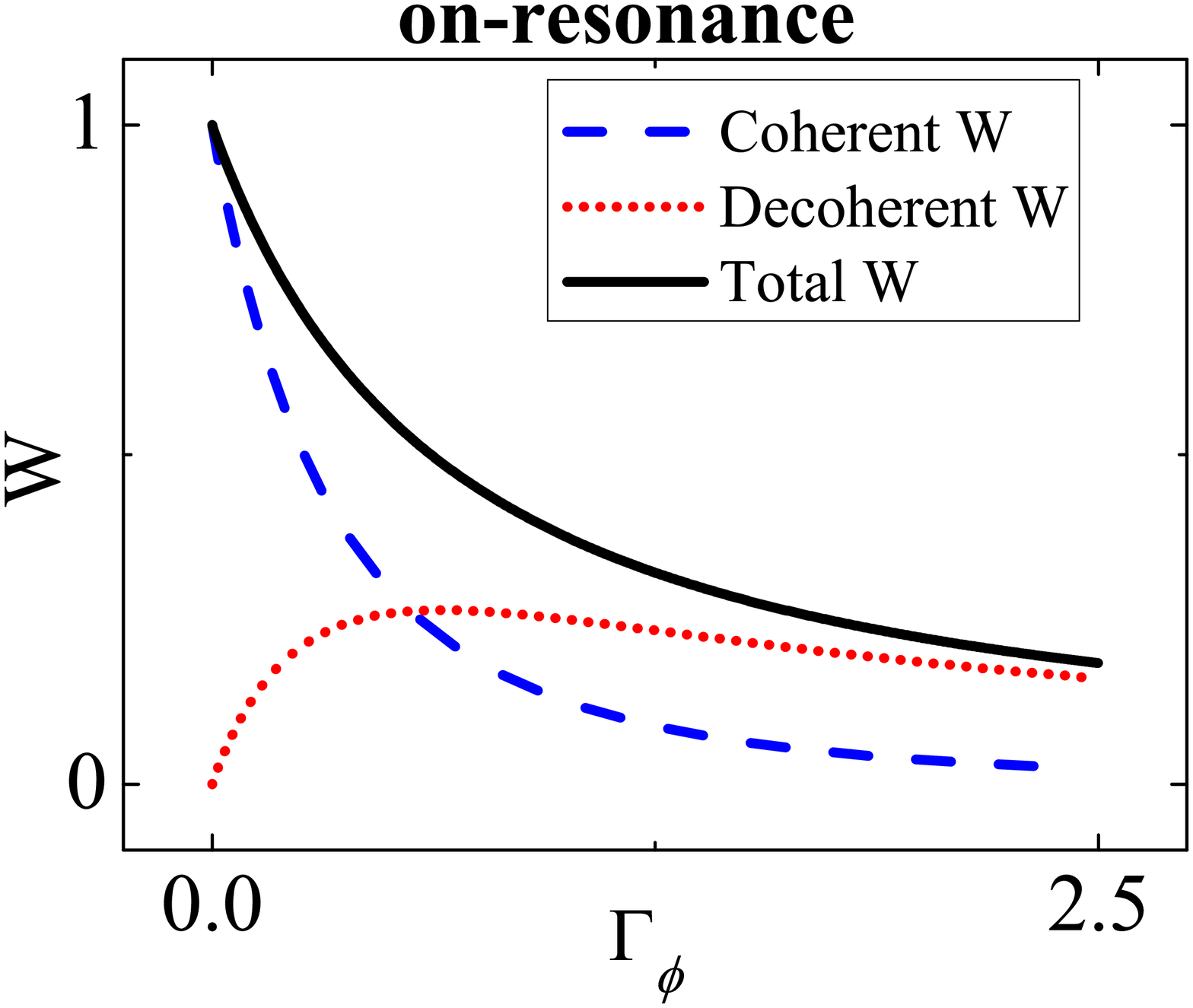}}
      \subfigure{\includegraphics[width=1.5 in,height=1.5 in,trim=0.5in 0.0in 0.1in 0.15in,clip=true]{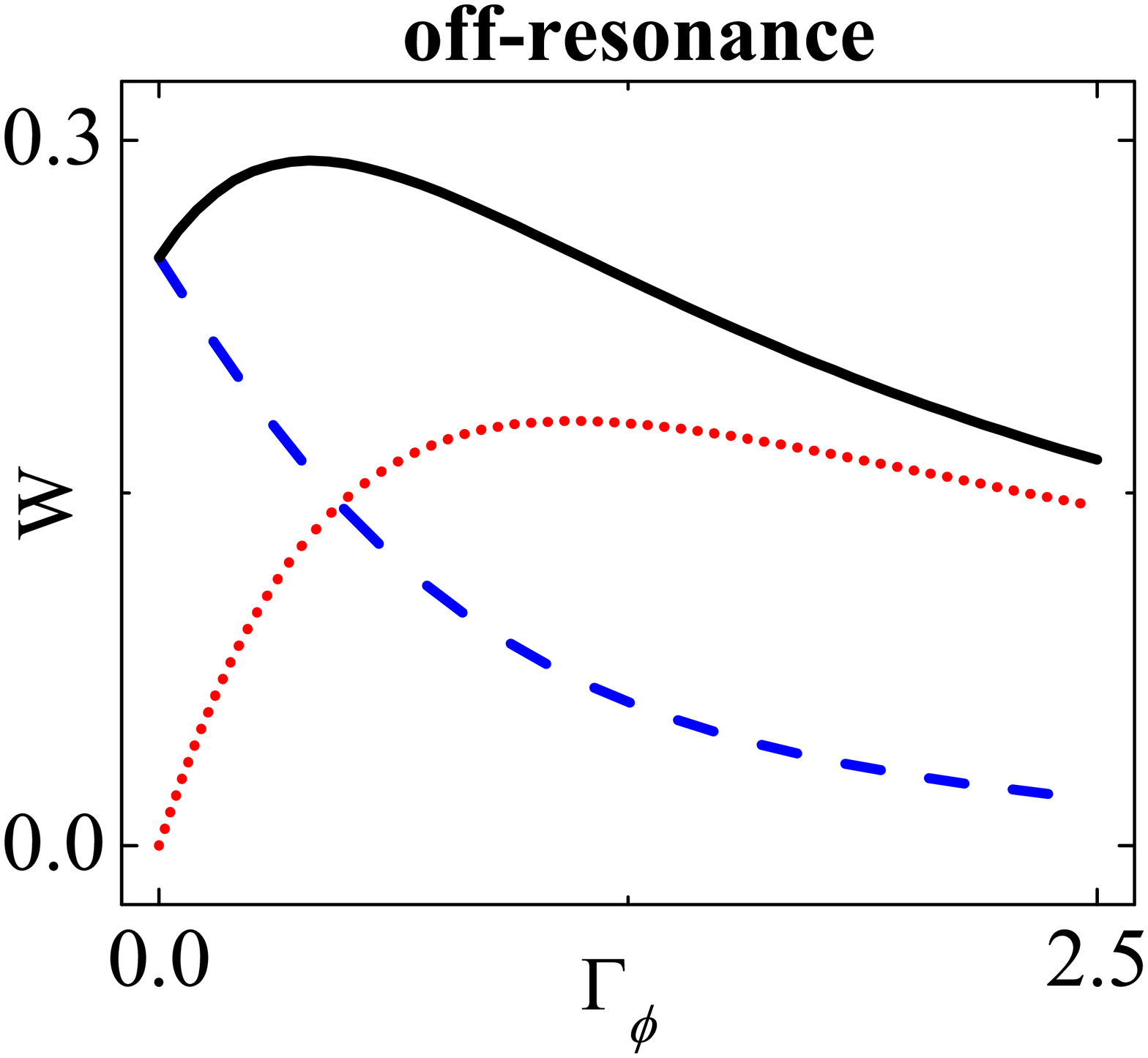}}      
    \end{center}
    \caption{ (Color online) \textbf{Upper Fig.-} Total work of the AQM as function of $(E_0-\varepsilon)$ and $\Gamma_\phi$, both expressed in units of $\Gamma_0$. The work is normalized to its maximum value in the figure. \textbf{Lower Figs.-} Work of the AQM for the ``on-'' and ``off-`` resonance conditions, $(E_0-\varepsilon)=0$ and $-1$ respectively. Different contributions are marked in the inset. In all figures, the  work is normalized to the maximum value of the total work.}
   \label{fig:work}
\end{figure}

\subsection{Efficiency of AQMs.}

The thermodynamic efficiency, $\eta _{TD}$, can be defined as the useful output power that can be  extracted from the system over the total input power. The useful output power is the work per cycle of the motor minus the energy lost due to friction, all divided by the period $\tau$, i.e. $Q^{}V / \tau-\int_0^\tau \gamma {\dot x}^2dt / \tau$. The total input power is $IV+Q^{}V/\tau$. Then,
\begin{equation}
\eta_{_\mathrm{TD}}=\frac{Q^{}-4 \pi^2 \gamma^*/\left(
\tau \texttt{V}\right) }{\bar{g} \tau \texttt{V} +Q}.  \label{eq:eficiency_TD}
\end{equation}
Here, we have introduced $I=\bar{g} V$ where $\bar{g}$ is the average conductance, and the corrected average friction coefficient $\gamma^*=d^2 \bar\gamma$ where $\bar{\gamma}$ is the average friction coefficient $\bar{\gamma}= \oint \gamma \dot{x}^2 dt / \oint \dot{x}^2 dt$. The dynamical constant $d$ essentially accounts for the deviations of $\dot x$ respect to its mean value, $d=\tau/(2\pi) \sqrt{\langle\dot{x}^2\rangle}$ where $\langle \dot{x}^2 \rangle=\oint \dot{x}^2 dt / \tau$.  

From Eq. \ref{eq:eficiency_TD} one can extract the minimum energy necessary for $\eta \neq 0$, which is the minimum energy necessary to start the motor's motion $Q V > 4\pi^2 \gamma^*/\tau $. Additionally one can also realize that there is an optimal value of $\tau$ that maximizes the efficiency,
\begin{equation}
\tau_{0}=\frac{4\pi^2 \gamma^*}{Q \texttt{V}}\left( 1+\sqrt{1+\frac{Q^{2}}{4 \pi^2 \gamma^* \bar{g}}} \right) , \label{eq:tau-optimo}
\end{equation}
This value can be used to find the optimal load that the motor can move or, given a load, the optimal voltage to be applied to maximize the efficiency. 
Note that, when the average conductance goes to zero, $\tau_{0}$ goes to infinity, which is consistent with the adiabatic limit of the efficiency proposed in Ref. \cite {AQM13}.

\section{Results.}

Just to illustrate our theory, we will consider a simple example, Fig. \ref{fig:scheme}, of a quantum dot that while rotating changes its resonant energy, $E(\theta)=E_{0}+\Delta E\cos (\theta+\theta_{0})$, and the coupling to one reservoir, $V_{R}+\Delta V\sin (\theta)$. The angular coordinate is $\theta$ and $\theta_0$ is a phase shift.
The couplings to the other reservoir $L$ and to the fictitious probe $\phi$ are assumed constant.
The system-environment interaction rate $2\Gamma_{\phi}/ \hbar$ is determined by the coupling to the fictitious probe.
The details of the solution of this example can be found in Appendix A. The main assumptions used are: 
$(1)$ The Landauer-B\"uttiker picture of non-interacting electrons is valid.
$(2)$ The interaction with the leads can be taken within the wide band limit (WBL).
$(3)$ The interaction between the dot and the movable part of the system is perturbative, i.e. the variations of $\Delta E$ and $\Delta V$ are small respect to $\Gamma_0$, the width of the dot's resonance without decoherence at $\theta=0$.
$(4)$ The terminal velocity of the rotor in the stationary regime is approximately constant, i.e. $d \approx 1$. We briefly discuss the conditions for the validity of this last in Appendix B.

\begin{figure}[ptb]
     \begin{center}
        \subfigure{\includegraphics[width=2.5in, trim=0.5in 0.1in 0.1in 0.3in, clip=true]{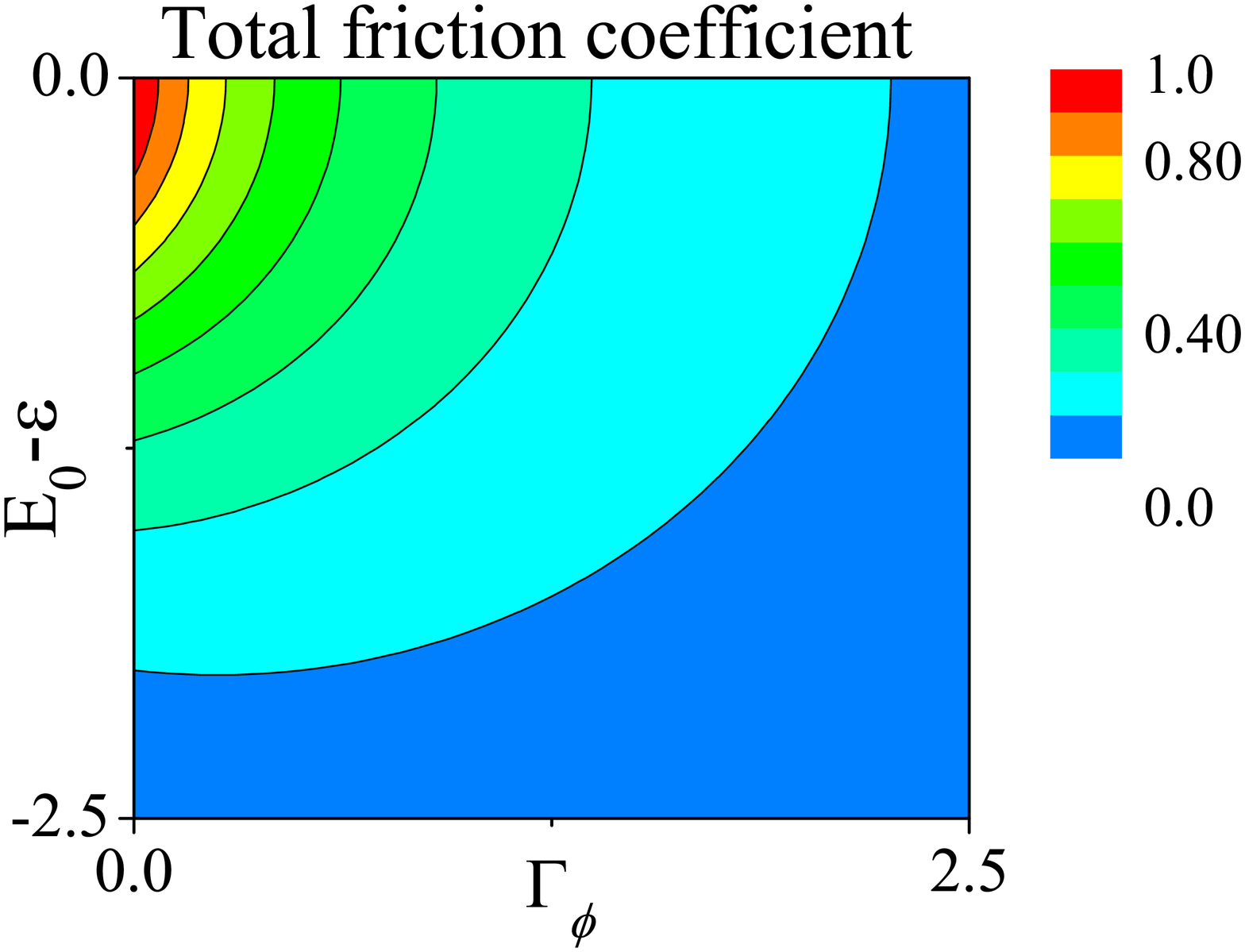}}        
                \\
        \subfigure{\includegraphics[width=1.5 in,height=1.5in, trim=0.5in 0.1in 0.1in 0.3in, clip=true]{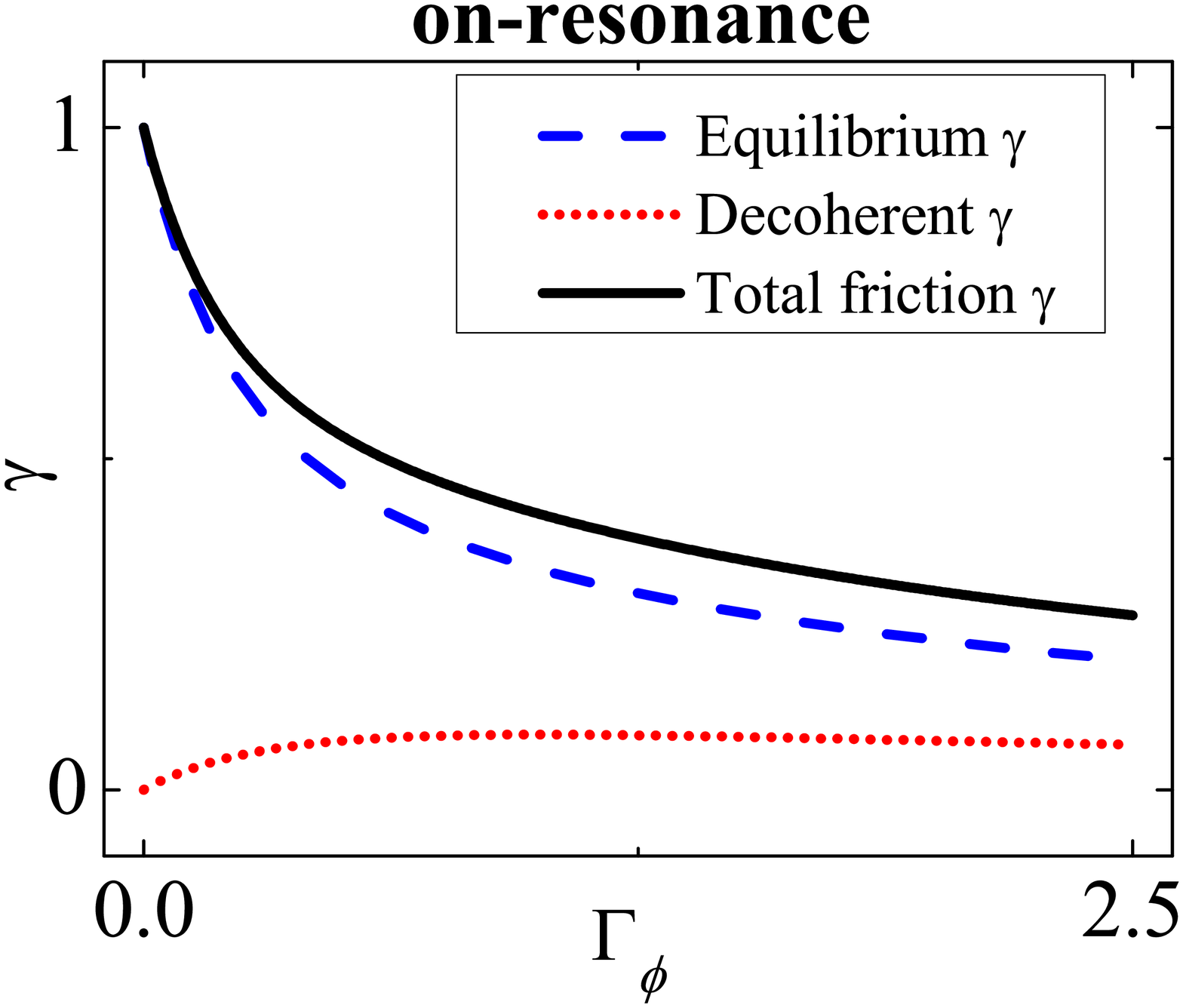}} 
        \subfigure{\includegraphics[width=1.5 in,height=1.5in, trim=0.5in 0.1in 0.1in 0.2in, clip=true]{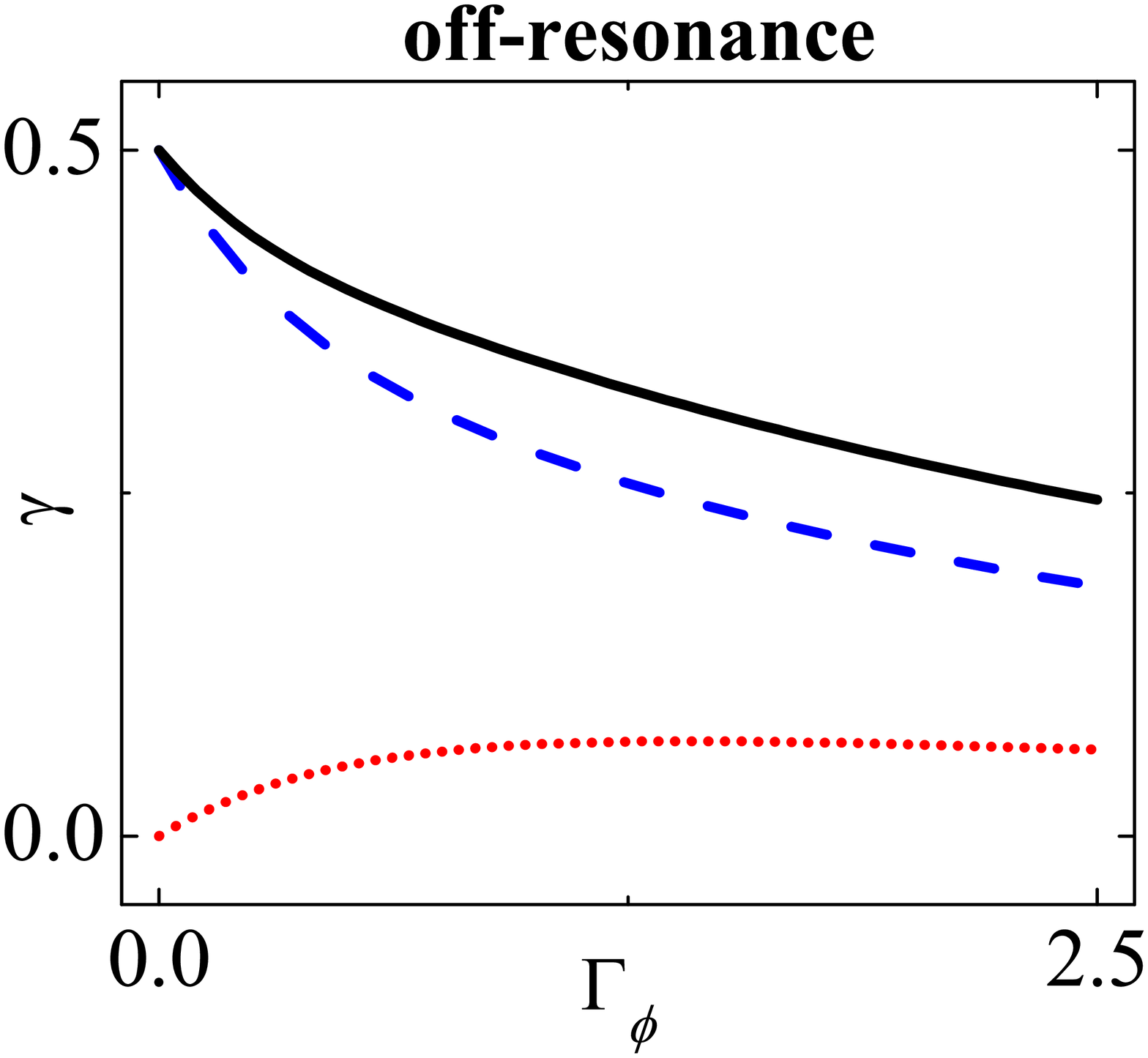}}
    \end{center}
    \caption{ (Color online) \textbf{Upper Fig.-}  Total friction coefficient normalized to its maximum value as function of $E_0$ and $\Gamma_\phi$, both expressed in units of $\Gamma_0$. 
    \textbf{Lower Figs.-} Friction coefficient of the AQM for the ``on-'' and ``off-`` resonance conditions, $(E_0-\varepsilon)=0$ and $-1$ respectively. Different contributions are marked in the inset. In all figures, the  values of the friction coefficients are normalized to the maximum value of the total friction coefficient.}
   \label{fig:friction}
\end{figure}
\begin{figure}
     \begin{center}
        \subfigure{
         \includegraphics[width=2.9 in, trim=0.0in 0.0in 0.0in 0.1in, clip=true]{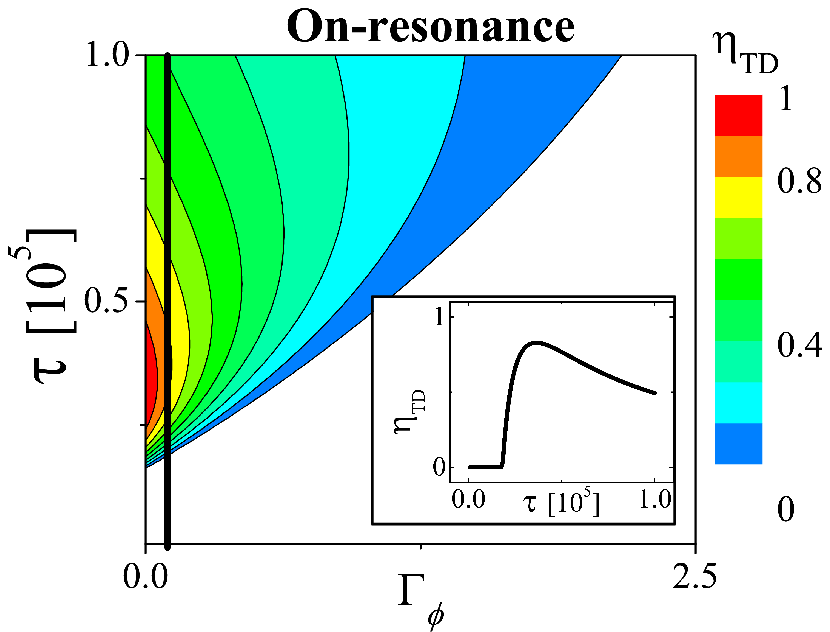}        }        
                \\
        \subfigure{
         \includegraphics[width=2.9 in, trim=0.0in 0.0in 0.0in 0.1in, clip=true]{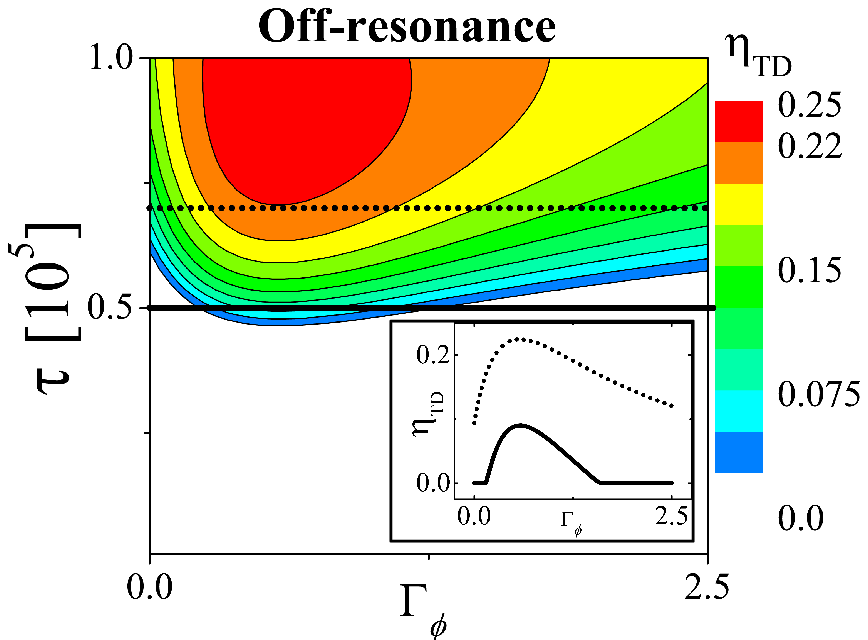}
        }      
    \end{center}
    \caption{ (Color online) Thermodynamic efficiency of the AQM shown in Fig. \ref{fig:scheme} as function of the decoherent rate $\Gamma_{\phi}$ (in units of $\Gamma_{0}$) and the period of the motor $\tau$ (in units of  $\hbar/2\Gamma_{0}$). \textbf{Upper Fig.-} on-resonance regime. \textbf{Lower Fig.-} off-resonance regime ($E_0-\varepsilon=-0.75 \Gamma_{0}$). The efficiency is normalized to its maximun value in all plots. Insets show the cuts marked in the main figures as continuous or dot lines.}
   \label{fig:efficiency}
\end{figure}
The work of the motor can be split into a coherent and a decoherent contributions. 
The behavior of $W$ is similar to that described in Ref. \cite{MkBt01} for the (decoherent) pumped charge. Details of the solution can be found in Appendix A, but the main results are shown in Fig. \ref{fig:work}.
In the on-resonance regime, i.e. when the Fermi energy $\varepsilon$ is  $\varepsilon \approx E_{0}$, the coherent contribution to the work is a monotonic decreasing function of  $\Gamma _{\phi}$, whereas the decoherent term increases with $\Gamma _{\phi }$ until it reaches a maximum value to decay afterward.
The total work is always a decreasing function of $\Gamma _{\phi}$. 
In the off-resonance regime, the coherent and the decoherent contributions behave qualitatively as in the on-resonance regime.
However, here the decoherent term dominates and, then, the total work presents a maximum for a finite value of $\Gamma _{\phi }$. This implies that in the off-resonance regime an adequate environment interaction can indeed maximize the work of the motor.

The friction coefficients $\bar{\gamma}^{s,eq}$ and $\bar{\gamma}^{\phi}$ present a similar behavior both in the on-resonance and in the off-resonance regimes as function of $\Gamma_{\phi}$, see Fig. \ref{fig:friction}.
$\bar{\gamma}^{s,eq}$ is a monotonically decreasing function of $\Gamma_{\phi}$ while $\bar{\gamma}^{\phi }$ presents a maximum.
The total friction coefficient $\gamma =\gamma^{s,eq}+\gamma ^{\phi }$ always decays with $\Gamma_{\phi}$,
both in the on-resonance and in the off-resonance regimes.
In principle, just by looking Eqs. \ref{eq:force} and \ref{eq:gammaPhi} one could (naively) expect that electronic
friction is increased due to the extra friction term. However, electronic 
friction is actually mitigated by decoherence. This is because decoherence diminishes the quantum fluctuations and consequently also the friction coefficient, due to the FDT. 

The effect of decoherence on the thermodynamic efficiency is complex due to a competition between 
the total work per cycle and the friction coefficient.
In fig. \ref{fig:efficiency} we plot the efficiency as function of $\Gamma _{\phi}$ and the period $\tau$.
As predicted by Eq. \ref{eq:tau-optimo}, there is a (decoherence dependent) optimal value of $\tau$ that maximizes the efficiency.
In this particular example the maximum efficiency is quite low, $1.6~10^{-5}$.
This is expectable considering the perturbative interaction between the rotor and the dot, which implies a small $Q$, and the high coupling between the dot and the reservoirs, which gives a high $I^{bias}$.
Beyond that, we find that, in the on-resonance regime, decoherence always diminishes the efficiency of the AQM.
On the contrary, in the off-resonance regime, there is a maximum in the efficiency for a finite value of $\Gamma _{\phi }$.
This shows that, surprisingly, efficiency can be increased due to a properly tunned interaction with the environment.
Indeed, there are certain values of $\tau$ where at a critical $\Gamma _{\phi}$ the efficiency switches from zero to finite values.
This implies that the AQM is indeed triggered by its interaction with environment.

\section{Conclusions.}

In conclusion, we developed the general theory of decoherent current induced forces and we applied
it to AQMs.
We showed that decoherence not only modifies the current induced forces but also the electronic or intrinsic dissipation mechanisms and its related force fluctuations.\cite{comment,comment2}
We proved that the theory is consistent with the fluctuation-dissipation theorem. Besides, we showed that the relation between the total work and the pumped charge per cycle, proved for coherent AQMs,\cite{AQM13} remains valid in presence of decoherence.
We exemplified our theory with a simple example of AQMs showing that even there the role of decoherence can be non trivial.
Indeed, we showed that decoherence can increase the efficiency of the AQM. This may allow the design of novel devices such as environmentally activated nanomotors.

\section{Acknowledgement.}

We would like to acknowledge F. von Oppen for useful discussions and comments. 
This work was supported by
CONICET, Secretar\'ia de Ciencia y Tecnolog\'ia de la Prov. de C\'ordoba, SECYT-UNC, ANPCyT and MINCyT.

\appendix

\section{Details of the calculation of the example of an AQM with decoherence.}

In this appendix we give the details of the calculations of the example presented in the main text, an AQM based on a peristaltic pump.
The energy level of this AQM depends on $\theta$, the rotational coordinate, as $E(\theta)=E_{0}+\Delta E\cos (\theta+\theta_{0} )$, where $\theta_0$ is a phase shift. The coupling to the reservoir $R$ is $V_{R}+\Delta V_{R}\sin (\theta)$.
We assumed a non-interacting electron picture.
All the solutions are obtained considering a perturbative interaction between electrons and the mechanical degree of freedom, i.e. $\Delta E$ and $\Delta V_{R}$ are small respect to $\Gamma_{0}$, the width of the dot's resonance without decoherence at $\theta=0$.
The chemical potential of the right reservoir is taken as the reference energy and we assumed a wide band approximation.
All observables are evaluated at low temperatures and in a linear response regime, where $e\texttt{V}\rightarrow 0$.
All this implies $\varepsilon=\mu_{R}$ in all the equations.

The effective Hamiltonian of the whole system, i.e., dot plus leads, reads
\begin{equation}
H(\theta)=E(\theta)+\Sigma _{L}+\Sigma _{R}(\theta)+\Sigma _{\phi},
\end{equation}
where $\Sigma_{i}$ is the self energy of the lead $i=L,R,\phi$.
We consider the leads within the wide band limit (WBL) where the self-energies $\Sigma$'s are pure imaginary quantities and are independent of the Fermi energy $\varepsilon$. Thus, 
$\Sigma _{L}=-\mathrm{i}\Gamma _{L}$, 
$\Sigma _{\phi }=-\mathrm{i}\Gamma _{\phi }$, and
$\Sigma _{R}(\theta)=-\mathrm{i}\Gamma _{R}(\theta)= -\mathrm{i}[V_{R}+\Delta V_{R} \sin (\theta)]^{2}/V_{_{BW}}$, where $4V_{_{BW}}$ is the bandwidth of the lead.
Here, the quantities $\Gamma _{L}$ and $\Gamma _{\phi}$ are constants since the couplings to their corresponding leads are assumed independent of $\theta$.
The retarded Green function, defined as $G_{0}=[\varepsilon-H(\theta)]^{-1}$, is
\begin{equation}
G_{0}=\frac{1}{\varepsilon- \left ( E(\theta)-\mathrm{i}\Gamma _{L}-\mathrm{i}\Gamma_{R}(\theta)-\mathrm{i}\Gamma _{\phi} \right)}. \label{eq:G0}
\end{equation}
Transmittances $T_{m,n}$, needed to calculate the observables from Eqs.  \ref{eq:QpumpXcycle} to \ref{eq:work}, can be obtained from
\begin{equation}
T_{m,n}=2\Gamma_{m}\left\vert G_{m,n}\right\vert ^{2}2\Gamma_{n},
\end{equation}
where $m \neq n$, and $m,n=L,R,\phi$. In this example, $G_{m,n}=G_{0}$, for all $m,n$. We can also obtain the density of states (DOS), $N(\varepsilon,\theta)=-(1/\pi) \mathrm{Im} (G_0)$, as
\begin{equation}
N(\varepsilon,\theta)=\frac{1}{\pi}\frac{\Gamma (\theta)}{(\varepsilon -E(\theta))^2+\Gamma (\theta)^{2}},
\end{equation}
where $\Gamma (\theta)=\Gamma _{L}+\Gamma _{\phi }+\Gamma _{R}(\theta)$.
The complete S-matrix can also be obtained from \ref {eq:G0}
\begin{equation}
\mathbb{S}=\mathbb{I}-2\mathrm{i}G_{0}
\begin{pmatrix}
\Gamma _{L} & \sqrt{\Gamma _{L}\Gamma _{R}(\theta)} & \sqrt{\Gamma _{L}\Gamma
_{\phi }} \\ 
\sqrt{\Gamma _{L}\Gamma _{R}(\theta)} & \Gamma _{R}(\theta) & \sqrt{\Gamma
_{R}(\theta)\Gamma _{\phi }} \\ 
\sqrt{\Gamma _{L}\Gamma _{\phi }} & \sqrt{\Gamma _{R}(\theta)\Gamma _{\phi }} & 
\Gamma _{\phi }
\end{pmatrix}.
\end{equation}
Thus, the matrix $\left[S^{\dagger }\frac{\mathrm{d}S}{\mathrm{d}\theta}\right ]$ gives
\begin{equation}
S^{\dagger }\frac{\mathrm{d}S}{\mathrm{d}\theta}=-2\mathrm{i}\left\vert
G_{0}\right\vert ^{2}\Lambda , \label{eq:S_dS_dx}
\end{equation}
where the elements of the operator $\Lambda$ are
\begin{eqnarray*}
\Lambda _{1,1} &=&\Gamma _{L}\frac{\mathrm{d}E(\theta)}{\mathrm{d}\theta}, \\
\Lambda _{1,2} &=&\sqrt{\Gamma _{L}\Gamma _{R}(\theta)}\frac{\mathrm{d}E(\theta)}{\mathrm{d}\theta}+\\
&&+\sqrt{\frac{\Gamma _{L}}{4\Gamma _{R}(\theta)}}\left( \varepsilon
-E(\theta)+\mathrm{i}\Gamma (\theta)\right) \frac{\mathrm{d}\Gamma _{R}(\theta)}{\mathrm{d}\theta}, \\
\Lambda _{1,3} &=&\sqrt{\Gamma _{L}\Gamma _{\phi }}\frac{\mathrm{d}E(\theta)}{
\mathrm{d}\theta}, \\
\Lambda _{2,2} &=&\Gamma _{R}(\theta)\frac{\mathrm{d}E(\theta)}{\mathrm{d}\theta}+\left(
\varepsilon -E(\theta)\right) \frac{\mathrm{d}\Gamma _{R}(\theta)}{\mathrm{d}\theta}, \\
\Lambda _{2,3} &=&\sqrt{\Gamma _{R}(\theta)\Gamma _{\phi }}\frac{\mathrm{d}E(\theta)}{
\mathrm{d}\theta}+  \\
&& +\sqrt{\frac{\Gamma _{\phi }}{4\Gamma _{R}(\theta)}}\left(
\varepsilon -E(\theta)+\mathrm{i}\Gamma(\theta)\right) \frac{\mathrm{d}\Gamma _{R}(\theta)}{
\mathrm{d}\theta}, \\
\Lambda _{3,3} &=&\Gamma _{\phi }\frac{\mathrm{d}E(\theta)}{\mathrm{d}\theta},
\end{eqnarray*}
with  $\Lambda _{m,n}=\Lambda _{n,m}^{\ast }$. 

At this point, we can evaluate all the physical quantities that are relevant to this problem.
Let us start with the emissivities of Eq. \ref{eq:emmissivity}.
At low temperatures, we can use $\partial f_{m}/\partial \varepsilon =-f_{m}(1-f_{m})/K_{B}T\simeq-\delta(\varepsilon-\mu_m)$. Thus, we have
\begin{eqnarray}
\frac{\mathrm{d}n_{m }}{\mathrm{d}\theta} &=&\frac{1}{2\pi \mathrm{i}}
\mathrm{Tr}\left\{ \Pi _{m }S^{\dagger }\frac{\mathrm{d}S}{\mathrm{d}\theta}\right\}
\\
&=&\frac{1}{2\pi \mathrm{i}}\left( S^{\dagger }\frac{\mathrm{d}S}{\mathrm{d}\theta
}\right) _{m ,m },
\end{eqnarray}
where $m =L,R,\phi $. Then
\begin{align}
\frac{\mathrm{d}n_{L(\phi )}}{\mathrm{d}\theta} & = -\frac{N(\varepsilon,\theta))}{\Gamma(\theta) }\frac{\mathrm{d}E(\theta)}{\mathrm{d}\theta}\Gamma _{L(\phi )}, \notag \\
\frac{\mathrm{d}n_{R}}{\mathrm{d}\theta} & = -\frac{1}{\pi }\frac{N(\varepsilon,\theta))}{\Gamma(\theta) } \notag \\
& \times \left[ \frac{\mathrm{d}E(\theta)}{\mathrm{d}\theta}\Gamma _{R}(\theta)+
\frac{\mathrm{d}\Gamma _{R}(\theta)}{\mathrm{d}\theta}\left( \varepsilon -E(\theta)\right)
\right]. \label{eq:emissivities}
\end{align}
We can insert these expressions into Eq. \ref{eq:QpumpXcycle} to obtain the pumped charge per cycle or the total work, note that $W =Q \texttt{V}$. The result can be split into a coherent and a decoherent contributions, $W=W^{coh}+W^{dec}$. The mathematical expression for $W^{coh}$ is exactly the same with or without decoherence (however, its value depends on $\Gamma_\phi$). Consistently, $W^{dec}$ is zero in a purely coherent case.

By evaluating Eq. \ref{eq:work} with Eq. \ref{eq:emissivities} and
using Green's theorem, \cite{Brower98} we obtain
\begin{equation}
W = -e\texttt{V} (\Omega_{coh}+\Omega_{dec})  \Delta E\Delta V_{R}\cos (\theta_{0}),
\end{equation}
where
\begin{eqnarray}
\Omega_{coh} &=& \frac{\Gamma _{L} V_{R}}{\Gamma_{T} V_{_{BW}}}
\left [ 4 \pi^{2} N^{2}(\varepsilon) \right ], \\
\Omega_{dec} &=& \frac{\Gamma _{\phi} \Gamma _{L} V_{R}}
{\Gamma _{0} \Gamma _{T} V_{_{BW}} } \left[ \frac{2 \pi N(\varepsilon)}{\Gamma _{0}} 
+4\pi^2 N^2(\varepsilon)\right].
\end{eqnarray}
Here, $\Gamma _{T}=\Gamma (\theta)|_{\Delta V_{R}=0,\Delta E=0}$ which does not depend on $\theta$, and $\Gamma _0=\Gamma _{L}+V_{R}^{2}/V_{_{BW}}$.
These equations are expressed at zero order, i.e., for small $\Delta E$ and $\Delta V_{R}$. Note that the factor $\Delta E \Delta V_{R} \cos(\theta_0)$ is the parametric area enclosed in a cycle of the parameter $\theta$.

The friction coefficient at equilibrium, $\gamma ^{s,eq}$, is
\begin{align}
\gamma^{s,eq} = \frac{1}{4}\sum\limits_{m ,n} & \int \frac{
\mathrm{d}\varepsilon }{2\pi }\frac{\partial }{\partial \varepsilon }\left(
f_{m }+f_{n}\right) \notag \\
& \times \mathrm{Tr}\left\{ \Pi _{m }S^{\dagger }
\frac{\mathrm{d}S}{\mathrm{d}\theta}\Pi _{n}S^{\dagger }\frac{
\mathrm{d}S}{\mathrm{d}\theta}\right\}.
\end{align}
Assuming a low temperature limit to evaluate the integral, yields
\begin{equation}
\bar{\gamma}^{s,eq} =\frac{\hbar }{8\pi^{2} }\oint \sum\limits_{m
,n}\left\vert \left( S^{\dagger }\frac{\mathrm{d}S}{\mathrm{d}\theta}\right) _{m ,n}  \right\vert ^{2}\mathrm{d}\theta . \label{eq:friction_coeff}
\end{equation}
The additional friction term, which is a direct consequence of decoherence on current induced forces, can be evaluated from Eq. \ref{eq:gammaPhi}, giving
\begin{equation}
\bar{\gamma}^{\phi } = \hbar \frac{\Gamma _{\phi }}{4\Gamma _{0}\Gamma _{T}}
N(\varepsilon)\Delta E^{2}.
\end{equation}
Notice that positivity of both coefficients are guaranteed.

\textit{List of parameters.} The parameters used in the whole work to perform the calculations are: 
$e\texttt{V} = 10^{-4}~V_{_{BW}}$,
$V_R = \sqrt{0.1}~V_{_{BW}}$,
$\Gamma_L = 0.1~V_{_{BW}}$,
$\Delta V_{R} =10^{-2} ~ V_R$,
$\Delta E = 2 ~10^{-3}~ V_{_{BW}}$, and
$\Gamma_0=\Gamma_L+V_R^2/V_{_{BW}} = 0.2~V_{_{BW}}$.
The maximum values of the efficiency, total work, and total friction coefficient used to
normalize the plots are  $1.6 ~ 10^{-5}$, $10^{-4}e\texttt{V}$, and $2.4 ~10^{-5}\hbar$ respectively.

\section{Limit of constant terminal velocity.}

At steady state, all the energy that is absorbed by the motor is completely dissipated by friction and thus, total energy is conserved
according to 
\begin{equation}
\int_{0}^{2\pi }\left[ F(\theta ) - F_{\mathrm{load}} -\gamma \dot{\theta}(\theta )\right] 
\mathrm{d}\theta =0.  \label{eq_Ener_cons}
\end{equation}
where $F_{\mathrm{load}}$ account for an external load to the system.
Here, we are assuming an average over steady state ensembles, so random forces can be neglected and we can use $\dot{\theta}(t)=\dot{\theta}(t+\tau)=\dot{\theta}(\theta)$.

From eq. \ref{eq_Ener_cons} and using the same definitions for $\bar{\gamma}$ and $\left\langle \dot{\theta}^{2}(t)\right\rangle$ as in the main text, we obtain,
\begin{equation}
\left\langle \dot{\theta}^{2}(t)\right\rangle =\frac{Q\texttt{V}-W_{\mathrm{load}}}{\bar{\gamma}}.
\end{equation}
where $W_{\mathrm{load}}$ is the work done by the external forces in one cycle.
Using this result, the mean kinetic energy becomes
\begin{equation}
\left\langle K \right\rangle=\frac{1}{2}I\left\langle \dot{\theta}^{2}(t)\right\rangle =\frac{1}{2}I\frac{(Q\texttt{V}-W_{\mathrm{load}})}{\bar{\gamma}}.
\end{equation}
where $I$ is moment of inertia of the rotor.

The change of the kinetic energy due to a rotation of the parameter $\theta $
is $\Delta K(\theta )=-\Delta U^{*}(\theta )$, where $U^{*}(\theta
)=-\int_{0}^{\theta }\left[ F(\theta ^{\prime })-F_{\mathrm{load}}-\gamma \dot{\theta}(\theta
^{\prime })\right] \mathrm{d}\theta ^{\prime }$ is a pseudo-potential defined only
at the steady state.
When the kinetic energy of the motor is much greater
than $\Delta U^{*}(\theta )$,
\begin{equation}
\frac{1}{2}I\frac{(Q\texttt{V}-W_{\mathrm{load}})}{\bar{\gamma}}\gg \Delta U^{*}(\theta ),\label{condition}
\end{equation}
then $\Delta K(\theta )/\left\langle K \right\rangle \rightarrow 0$ and, thus, the rotational velocity of the system becomes insensitive to $\theta$.
Therefore, $\tau =\int_{0}^{2\pi } \mathrm{d}\theta /\dot{\theta}(\theta )\approx 2\pi/\dot{\theta}$ and hence the dynamical factor $d=\frac{\tau }{2\pi }\sqrt{\left\langle \dot{\theta}^{2}(t)\right\rangle }\approx 1$.
Note that the terminal velocity is independent of $I$. Therefore, the condition given in Eq. \ref{condition} are always valid in the limit of macroscopic rotors, rotors with a sufficiently large moment of inertia.

\bibliography{./AQM}

\begin{thebibliography}{24}
\bibitem [{Coh()}]{Cohen}%
  \BibitemOpen
  \href@noop {} {\bibinfo  {journal} {Note that we can identify $F$ and $I$
  with the generalized forces associated with the generalized fluxes $\dot
  \theta$ and $\dot \Phi$, where $\Phi$ is the magnetic flux related to
  $\texttt{V}$ by $\texttt{V}=\dot \Phi$. See also, D. Cohen, Phys. Rev. B,
  \textbf{68}, 201303R (2003)}\ }\BibitemShut {NoStop}%
\bibitem [{com({\natexlab{a}})}]{comment}%
  \BibitemOpen
\bibfield  {journal} {  }\href@noop {} {\bibfield  {journal} {\bibinfo
  {journal} {Electron backscattering is the origin of current induced forces.
  Our model, being local, does not conserve electron momentum. This will
  naturally induce an additional contribution to the fluctuations of the force
  and consequently an additional friction process (as described by
  fluctuation-dissipation theorem). Non conservation of momentum is not a
  drawback of the model but indeed a general property of decoherence processes.
  From position-momentum uncertainty, it is clear that strict momentum
  conservation is not possible after decoherent events acting on finite
  systems}\ } ({\natexlab{a}})}\BibitemShut {NoStop}%
\end{thebibliography}

\end{document}